# Microwave derived *monoclinic* $Ba_{1-x}Sn_xNb_2O_6$ materials as an alternative of ITO

Ashish Joshi[a], Vaibhav Shrivastava[b,†], and Anurag Pritam[b]

Microwave synthesis was optimized for preparing novel monoclinic Tin-doped Barium Niobate ceramics ($Ba_{1-x}Sn_xNb_2O_6$; x= 0.0, 0.01, 0.05, 0.1, 0.2, 0.3) BSN. The intensity of monoclinic phase formation was observed to decrease on increasing tin as dopant indicating decreased crystallinity. Strained crystalline phase was observed in undoped sample that became severe on doping tin. Monoclinic metal alloy $Sn_2O_3$ formation was confirmed on increasing tin doping beyond 5%. Electronic configuration of tin (II) oxide supports the local site-wise monoclinic disorder in crystal structure due to sterically active lone pair. Such a disorder arranges increased number of degenerate energy states and reduce effective energy gap between conduction band minimum and valence band maximum. All BSN compositions were investigated for monoclinic phase stabilization, ultra-violet absorption, dielectric response, raman modes and type of carrier concentration along with hall resistivity. All measurements possessed inflexion corresponding to 5 atomic % tin doping indicating successful site substitution and estimated $Sn_2O_3$ metal formation beyond this. The values of reducing optical energy band gap, transparency to visible spectrum and hall resistivity indicated utility of these materials as a substitute transparent conducting oxide (TCO) for well-known indium tin oxide (ITO).

## I. INTRODUCTION

Tungsten bronze (TB) structure based polar dielectrics form a very important class of non-toxic materials offering competent structural, electrical, optical, photocatalytic and electrocaloric response compared to corresponding toxic perovskite oxides like lead magnesium niobate (PMN), lead zirconium niobate (PZN), lead zirconium titanate (PZT), lanthanum-modified lead zirconate titanate (PLZT), lead scandium niobate (PSN) etc[1-3]. Many TB structures offer variety of applications emerging from $NbO_6$ octahedral distortions and phase transition between different polymorphs. Carruthers and Grasso *et al.*[4] discussed ternary room temperature phase relationship

---

[†]Author for correspondence: vaibhav.shrivastava@snu.edu.in

[a]Department of Applied Physics, Delhi Technological University, Delhi-110042, India

[b]Dielectric Laboratory, Department of Physics, Shiv Nadar University, G.B.Nagar-201314, India

between BaO-SrO-Nb$_2$O$_5$ oxides. The members of mixed matrix Ba$_{1-x}$Sr$_x$Nb$_2$O$_6$, barium niobate BN (BaNb$_2$O$_6$) and Strontium Niobate (SrNb$_2$O$_6$) have not been discussed in detail for electro-optic coefficients, pyroelectric coefficients, and ferroelectric properties[5-7]. Current work was planned to utilize high transparency of barium-based dielectrics and instill high electric conductivity simultaneously to offer competitive alternative of Indium Tin Oxide, popular transparent conducting electrode material[8-10]. Indium tin oxide is an exhaustively used transparent conducting oxide (TCO) due to its high optical transparency (~85%) and electrical resistivity (~10$^{-4}$ Ω-cm.) and optical energy band gaps (~3.8eV). The thin film geometries of ITO are highly used in solar cells, touch screens, flat panel displays etc. Further, ITO being an n-type degenerate semiconductor was also proposed as photocathode material for dye-sensitized solar cell structure[10].

Taking versatility of applications offered by ITO, our choice was to start with polymorph barium niobate BN (BaNb$_2$O$_6$) whose phase transition studies makes it further interesting optimize for selective phase for diverse applications. However, the studies on its structural, optical and dielectric properties are sparse and need to be reported. Mixed tungsten-bronze SBN system was reported mostly in orthorhombic form instead tetragonal[6, 7]. In these structures, cationic site A1 (tetragonal site) is proposed to be occupied by Sr$^{2+}$ whereas A2-sites (pentagonal) can be occupied by both Sr$^{2+}$ and Ba$^{2+}$ in mixed matrix[5]. In BN, since there is no strontium thus A2-sites are expected to be occupied by barium ions. In the current work, microwave synthesis is optimized to prepare BN compound in monoclinic phase and further distort the unit cell by doping litharge structure type SnO for generating local disorder, hence, reduced crystallinity. The target is to introduce tin (Sn$^{2+}$) onto barium (Ba$^{2+}$)-sites to prepare Ba$_{1-x}$Sn$_x$Nb$_2$O$_6$ (BSN) composition series ($x$ = 0.0, 0.01, 0.05, 0.1, 0.2, 0.3). This proposed substitution is expected to impart lateral covalent π-π overlapping to increase electron density between two ionic positions of Ba-O bond. High electron density was expected to be achieved by inducing degeneracy in undoped BN structure on introducing tin. Thorough mechanical milling followed by homogenous microwave heating was expected to generate special *nano* dimension heat sinks on tin-sites in metal alloy form. Conducting metals possessing short skin depth re-radiate absorbed microwaves to the surrounding and generate local high temperature regions. On the other hand, atomic sites exhibiting prominent dielectric character, utilize absorbed microwave to resonate and generate controlled temperature regions by friction. The prepared compositions were further characterized for expected changes in crystalline phase formation, particle size variation, optical energy band gap, raman modes of vibration, *ac* conductance in correlation with rating of covalent character/ionic character in undoped BN and hall activity.

## II. EXPERIMENTAL PROCEDURE

Each stoichiometric composition $Ba_{1-x}Sn_xNb_2O_6$ (BSN) was prepared by taking high purity (≥99.9%) $BaCO_3$, $Nb_2O_5$ and $SnO$ powders (all from Sigma Aldrich Inc. USA). The mixed powders were grounded for three hours in covered electrical agate pulverizer in intermediate stages of spatula mixing. The mixtures were calcined at 1050°C for 3 hours using a microwave furnace operating at 2.45 GHz with deliverable power of 2.2 kW in automatic magnetron changing mode. The SiC susceptor cavities were used for concentrating microwave power in a region of 60X60X60mm. The heating (3°C per minute) and cooling (2°C per minute) profiles were monitored through Eurotherm 2416 controlled AST infrared pyro sensor. The calcined powders were uniaxially pressed into disc-shaped pellets to minimize density gradient by applying a pressure of 700MPa. The densification of pellets was carried via microwave sintering at 1050°C for 1 hour for removal of PVA added as an organic binder. X-ray diffractograms were recorded using Bruker D8 Advance model in coupled ($2\theta$-$\theta$) mode using Cu$K\alpha$ radiation of wavelength 1.54406Å in the range of angle $2\theta$ from 10° to 80°. Transmission Electron Microscopy (TEM) was engaged to analyze the grain growth and to compare particle size with crystallite size as estimated using Scherrer's expression. These images were captured using LIBRA 200FE high resolution TEM with information limit of 0.13 nm equipped with EDS, which operated at 200kV. Optical energy band gap for each modified BSN composition was calculated using *Tauc* plots. These plots were derived using UV-Vis diffuse absorption spectra measured in an integral sphere mode using spectrophotometer SHIMADZU UV-Vis 3700. For reflecting incident radiations $BaSO_4$ was used as a medium co-mixed with each modified BSN composition. The vibrational characteristics of all present chemical bonds in BSN compositions were studied using Fourier transform infrared (FTIR) spectroscopy. These FTIR spectra were recorded on FTIR Spectrophotometer of Thermo Fisher Scientific make and model Nicolet iS5 in the range of 500 and 4500 $cm^{-1}$. For recording this spectrum, each BSN composition was mixed with *KBr* powder and infrared light was shone on the prepared pellet. The pellets of prepared BSN composition were coated using silver and gold for the dielectric measurements. Dielectric data was collected using high frequency LCR meter NF2376 from NF Corporation, Japan in the frequency range 20 to 2MHz at temperatures from 25 to 575 °C.

## III. RESULTS AND DISCUSSION

**1. Crystal Structure and Morphology Analysis**

Perfect monoclinic phase formation is confirmed in all microwave prepared BSN samples using JCPDS 01-077-541, Fig.1. All the x-ray peaks are well indexed for monoclinic phase of polymorph $BaNb_2O_6$, also this phase remains preserved overall on tin doping. Most reports have a mention about orthorhombic phase of strontium doped $BaNb_2O_6$ generally admixed with tetragonal phase[9,10] yielding crystalline planes diffused within each other along oblique axes of orthorhombic and tetragonal symmetries. A few unindexed new peaks (110)/(101) prominently emerge between 25-26.9° and 33.1-34.3° on increasing tin doping after 5%. These peaks belong to tin in oxide

forms, $Sn_2O_3$ (JCPDS 00-025-1259), SnO (JCPDS 00-006-0395) and $SnO_2$ (JCPDS 00-088-0287) in descending order of their peak matching percentage respectively. These fix Bragg angle peaks support metal alloy formation embedded in dielectric BSN unit cell structure in preferred orientation. Minor increase in x-ray intensities of these peaks is crucial in concluding the pivotal role of microwave heating followed after thorough milling of mixtures. The metal clusters of tin are expected to receive maximum microwave power in atomic network and melt *before* barium carbonate to flow randomly. This random flow in eutectic form presumably fills the vacant sites of barium to saturate coordination number. Current monoclinic structure is known to be natural tungsten bronze type and typically useful in photovoltaic applications[11]. Nearly undetectable intensity change in x-ray diffractograms for samples up to x=0.05 is an indicative of tin successfully substituting barium on respective sites.

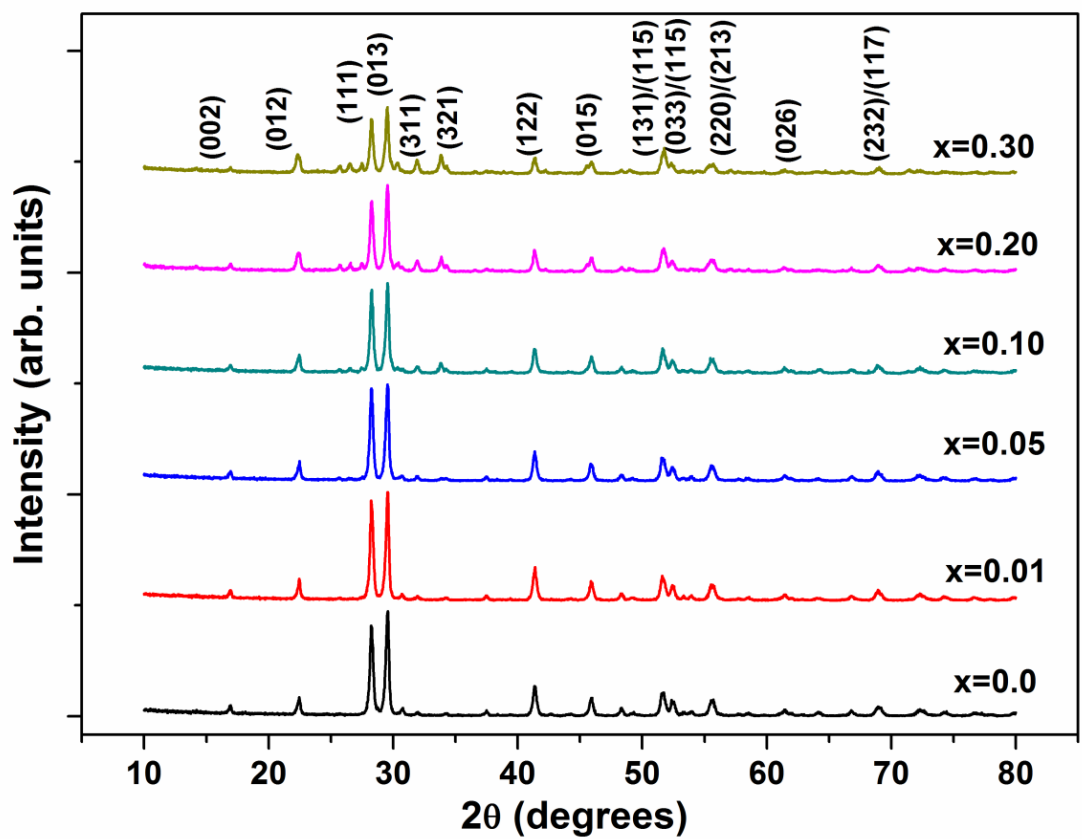

**Fig. 1. X-Ray diffractograms of $Ba_{1-x}Sn_xNb_2O_6$ composition pellets**

The introduction of tin beyond 5 % causes monoclinic metal alloy $Sn_2O_3$ formation in $BaNb_2O_6$ in addition to dominant monoclinic phase. The generated distortions are further investigated through calculations of distortion parameters like tetragonal strain and orthorhombic distortion. The lattice parameters are calculated using indexed peaks (012), (111) and (013) and used for computing distortion parameters. The values of lattice parameters (*a:*3.957Å, *b:*6.063Å, *c:*10.453Å) calculated are in accordance with those recommended for monoclinic BN. The variation in the values of lattice parameters *a*, *b* and *c* exhibits inflexion around 5% tin doping indicating exact site substitution without alloy formation till 5% doping. Tin (Sn: $Kr,4d^{10},5s^2,5p^2$) as dopant for barium (Ba: $Xe,6s^2$) promotes the formation of covalent bonds site wise compared to ionic interaction between barium and oxygen. Many reports have a mention about measure of increased covalence in the crystal structure in terms of tetragonality (c/a), the same is exercised in current work[12, 13]. An increase in tetragonal strain values (c/a) for initial doping up to 5% indicates about expected increase in covalence nature of chemical bonds due to c-axis oriented lateral π-π overlap. This finally results into lattice relaxation in *a-b* plane thereby decreasing orthorhombic distortion. Crystallite size (*t*) was estimated using the Scherrer's formula for the high intensity peaks including the maximum one: $t = \frac{k\lambda}{\beta cos\theta}$, where, $\lambda$ is the wavelength of x-ray used, $\theta$ is the position of the maximum intensity peak, $\beta$ is full width at half maxima (FWHM) of maximum intensity peak and *k*=0.9 is the Scherrer's constant. The crystallite size ranges from 27.1 nm to 29.5 nm and follows an inverse relationship with the tetragonal strain leaving sample x=0.01.

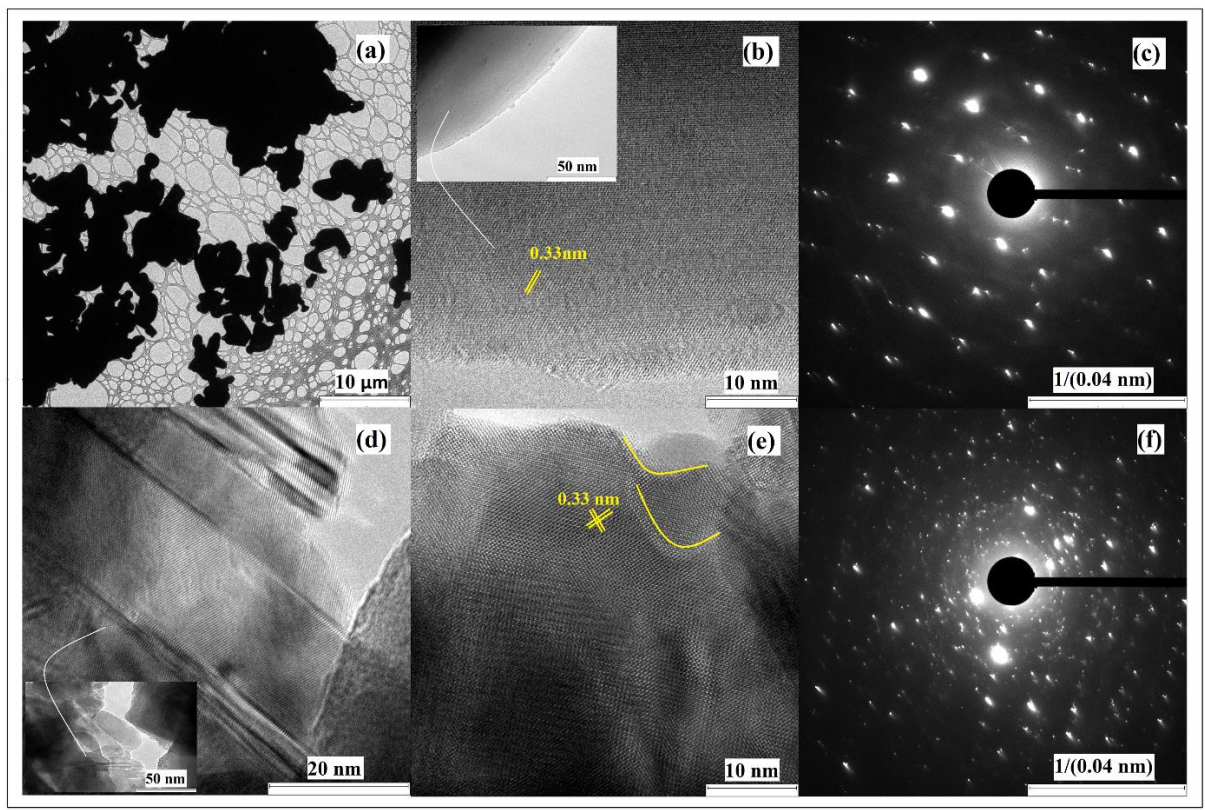

**Fig. 2. TEM images of Ba$_{1-x}$Sn$_x$Nb$_2$O$_6$ grains [(a)-(c) for x=0.00 and (d)-(f) for x=0.30]**

In current work, premeditated interaction of microwaves is executed with tin metal alloys regions in doped BSN compositions. Beyond 5% doping, tin is observed to follow alloy formation, which is behaving finally as metal in dielectric network to attract maximum fed microwave power[14]. Two typical samples (undoped BSN and 30% doped BSN) were taken from the crushed pellet powder to investigate the microstructure by transmission electron microscopy (TEM), showing the formation of nanocrystals. The microwave prepared BSN compositions contained bigger particles of the order of 1μm along with a few regions of *nano* particles [see Fig.2(a)]. The micron-sized crystals are showing long range order as the one shown in Fig.2(b), confirming low lattice strain. The diffraction spots observed in selected area electron diffraction (SAED) regime [Fig.2(c)] indicate the formation of strained crystalline network in undoped BaNb$_2$O$_6$ sample. This strain becomes severe in maximum doped BSN sample; showing clear interweaved zones indicated by arrows [see Fig.2(d)]. Further inspection of Figs.2(e) & (f) indicate the development of polycrystalline phase, while the grain boundaries are clearly visible in Fig.2(e)[15]. The local tin alloy regimes melted before barium carbonate and flow in eutectic form to channelize the quick preparation of monoclinic $BaNb_2O_6$ structure [see curved surfaces of grains in Fig.2(e)]. The distribution of energy dispersive spectrometry (EDS) peaks confirm presence of very close atomic stoichiometry as planned with no other elements present. EDS results indicate an increase in mass density of $Ba_{0.7}Sn_{0.3}Nb_2O_6$ matrix from 5.71 (for $BaNb_2O_6$) to 6.07 $g/cm^3$ on increasing tin by 30 atomic % (x=0.30). Tin possessing less atomic weight than barium should have caused decrease in mass density. This provides an indication of tin occupying sites other barium on high doping. The other available sites of 6-coordinated niobium favor tin occupancy; this was investigated further using spectroscopic methods.

## 2. Spectroscopic investigations

Fig.3 shows the optical absorption spectrum for investigated materials. The BSN compositions show major absorption in *uv* region starting from 250*nm* and concludes at 410*nm*. Such a broad absorption region is solely due to intrinsic transition of semiconductors and less likely due to defect energy states. The absorption peak broadens on increasing tin up to 10% and decreases thereafter. The broadening is investigated using full width at half maximum (FWHM), as an ancillary indication of grown semiconducting behavior of dielectric BSN compositions, Table-1. The absorption edge in all BSN compositions is found in the UV-Vis region from 380- 415*nm* indicating about energy gap between valence band maximum and conduction band minimum typically around 2.98-3.26eV, better than ITO[8,16,17]. The samples up to 5% doping show absorption edges shifting towards blue wavelength whereas the ones with higher doping exhibited red shift. The absorption values were below 0.1 for all samples, which is an essential requirement for TCO.

The samples with tin doping up to 5% possessed absorption lower than undoped BN that increased on tin doping beyond 5 atomic% justifying optimal tin substitution. Almost no change in the wavelength value for maximum absorption intensity supported typical strength of monoclinic phase of BN towards photo absorption after doping. The variation of absorption intensity is directly related to the crystallite size of the BSN samples. Samples with big crystallites absorb more UV radiation than those with small crystallites. Therefore, the response of absorption intensities possessed by samples x=0.05 and x=0.30 can be understood self. All BSN samples show a direct band gap transition and the calculated band gap energy values are listed in Table-1. A decrease in optical energy band gap on tin doping is due to significant increase in number of energy states compared to undoped BN. Additional transmittance measurements for a few samples; confirm the transparent response of BSN compositions to visible spectrum starting from 415*nm* because of transparent nature of constituent reduced oxides BaO, SnO and $Nb_2O_5$ [18-20].

**Table-1 Optical energy band gap values and FWHM for UV absorption in BSN samples**

| Composition (x) | FWHM (nm) | $E_g$ (eV) |
|---|---|---|
| 0.00 | 112.433 | 3.51 |
| 0.01 | 113.133 | 3.45 |
| 0.05 | 141.308 | 3.47 |
| 0.10 | 182.838 | 3.12 |
| 0.20 | 170.207 | 3.01 |
| 0.30 | 150.655 | 3.22 |

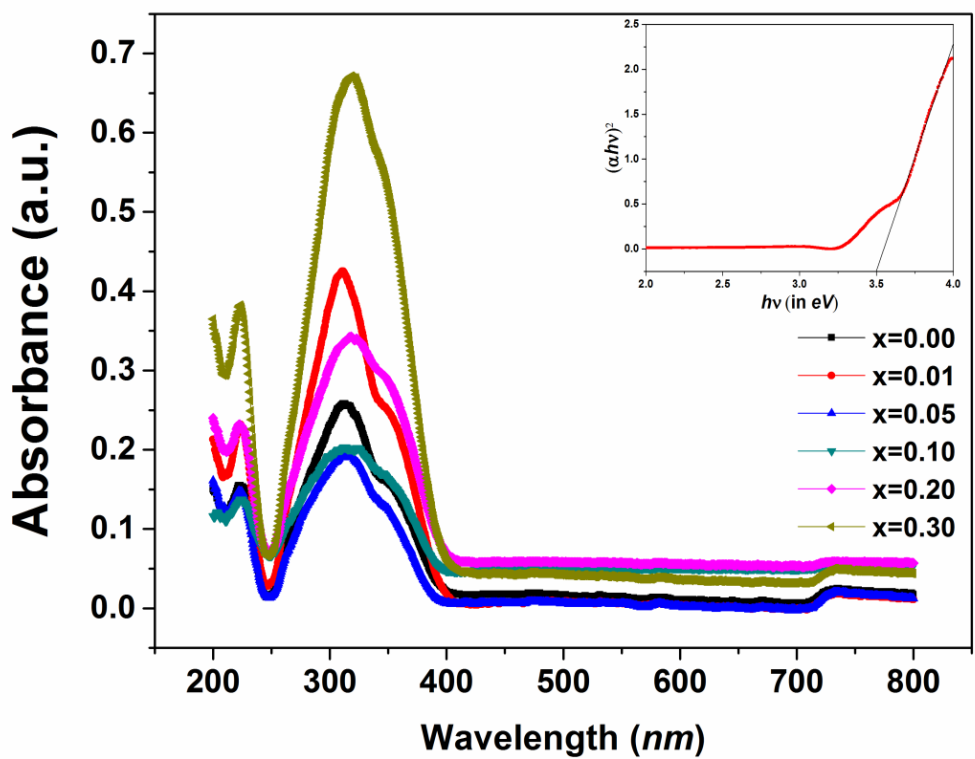

**Fig. 3. UV-Visible absorption spectra of $Ba_{1-x}Sn_xNb_2O_6$ compositions**

Tin induced structural distortions in BSN compositions are further investigated using micro-raman spectroscopy, Fig.4. It is an effective method for rating the structural perfectness of an atomic framework. The distorted framework possess additional low energy coupled vibrations usually detectable as scattering losses. Therefore, raman scattering is the best tool for detecting these coupled vibrations emerging due to inelastic scattering of laser photons after interaction with electrons in atoms at different atomic sites. In addition, active raman modes in any structure correspond to number of atoms per unit cell in that structure. The obtained raman spectra of BSN samples are in excellent agreement with the work of Repelin *et al* [21], who concluded about tilting of corner shared $NbO_6$ octahedrons generating typical orthorhombic unit cell structure that can further distort to generate monoclinic system. Fifteen active raman modes are present in undoped BN, confirming the 15-folded A2 site occupancy by barium in BN, Table-2. There is no shift in all peak positions except tin-alloy induced growth and splitting of peaks 2 and 13. Thus, parent monoclinic phase of BN remains unaltered except diffusion of additional monoclinic phase of $Sn_2O_3$ at 10% doping and above[22]. Tin oxide (SnO) occupies less polar and more *c*-axis strained *tetragonal* litharge structure[23]. Hence, tin doping infuses a decrease in *phonon driven intrinsic* *polarizability* per unit cell, therefore, all doped BSN compositions show decrease in intensity of all raman modes. Again, an increase in tetragonality (*c/a*) and no splitting of raman peaks can be correlated for tin doping up to 5% as estimated in previous sections. Beyond 5% doping, tin atoms not only occupy barium sites corresponding to unaltered raman peak positions rather generate local asymmetry to produce coupled vibrational modes due to monoclinic $Sn_2O_3$ formation as local distortion. These coupled modes appear as splitting of band-13 in present work. This is also confirmed by investigating full width at half maxima (FWHM) for vibration bands 1, 3, 5, 11 and 13. As tin is introduced in the BSN system (sample x=0.01), FWHM increases hence, raman bands become wider to adjust peak intensity to lower value. This further introduces a few more new peaks and shoulder bands starting from BSN composition x=0.1 in low frequency as well as in high frequency region. In x=0.2 and x=0.3 samples, the low frequency shoulder peaks are diminished while the shoulder peaks at high frequency bands become more pronounced.

**Table 2. Raman bands in $Ba_{1-x}Sn_xNb_2O_6$ compositions**

| Band | Frequency (cm$^{-1}$) | Band | Frequency (cm$^{-1}$) |
|---|---|---|---|
| 1 | 120 vs | 9 | 376 w |
| 2 | 146 w (for BN) + s (for $Sn_2O_3$) | 10 | 490 w |
| 3 | 181 s | 11 | 556 vs |
| 4 | 199 w | 12 | 627 w |
| 5 | 227 s | 13 | 707 vs (for BN) + s (for $Sn_2O_3$) |
| 6 | 278 m | 14 | 771 w |
| 7 | 303 m | 15 | 841 w |
| 8 | 360 w | | |

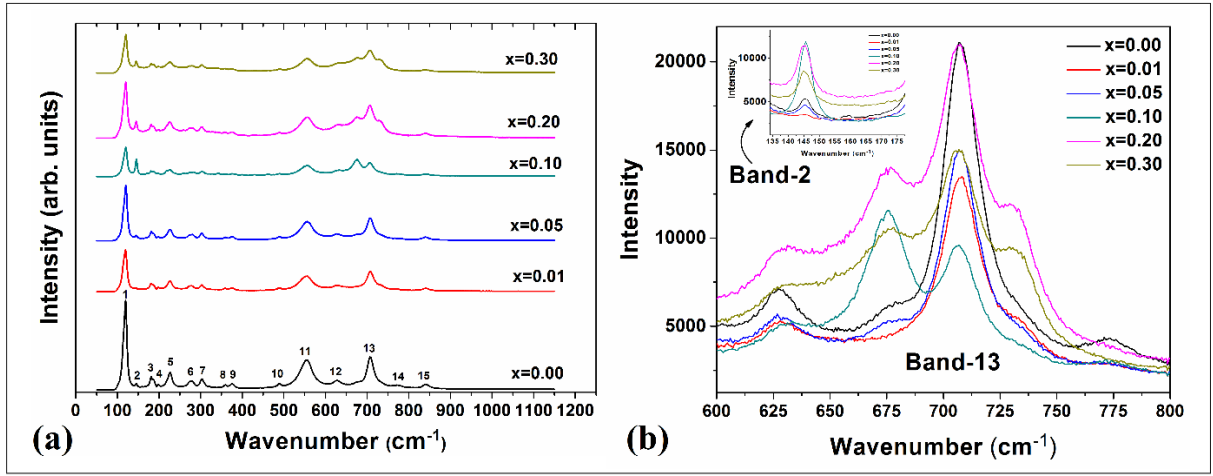

**Fig. 4. Raman Spectra of BSN compositions**

To study the intrinsic contributions to real permittivity of BSN compositions, which is normally observable at high frequencies (~$10^{12}$ Hz and beyond), FTIR spectroscopy is employed. Various chemical bonds present in a typical stochiometrically prepared composition vibrate after absorbing infrared radiation corresponding to typical expression given by,

$$\bar{v} = \frac{1}{2\pi c}\sqrt{\frac{k}{\mu}}$$

where $\bar{v}$ is the wavenumber (cm$^{-1}$), $c$ is the velocity of light (cm./sec.), $k$-is the force constant (dynes/cm.), and $\mu$ is the reduced mass (gms.). The results of FTIR spectra of the BSN samples are in exact correlation with raman modes appeared from 500 to 1000 cm$^{-1}$, Fig.5. Observed IR absorption bands are compared with the data already reported[24-26] for Ba-O, Sn-O and Nb-O bonds. The absorption bands between 500 cm$^{-1}$ and 600 cm$^{-1}$ are assigned to antisymmetric stretching vibrations of Nb-O bonds in NbO$_6$ octahedra. The absorption bands at 600 cm$^{-1}$ and 630 cm$^{-1}$ are corresponding to Ba-O and Sn-O bond antisymmetric stretching vibrations respectively. It can be seen that on increase tin doping the peak intensity corresponding to Ba-O bond decreases and the same for Sn-O bond increases. This intensity change confirms the successful replacement of barium ions by tin ions onto 'A2' sites. The metal-oxygen bonds are observed to resonate at energies corresponding to wavenumbers lower than 1000 cm$^{-1}$ range. While the vibration bands observed beyond 1000cm$^{-1}$ are corresponding to the response of carbon assisted functional groups to absorbed radiation. The absorption band at 1217 cm$^{-1}$ is attributed to C-O stretching vibrations, which could be either be due to residual carbonates in composition or because of adsorption of CO/CO$_2$ from the atmosphere. The presence of the absorption bands at 1363 cm$^{-1}$ and 1445 cm$^{-1}$ verify the presence of BaCO$_3$ minor traces mostly present on open powder surfaces due to CO/CO$_2$ adsorption during FTIR measurement. All expected absorption bands are observed in the investigated BSN samples. Absorption bands at 1730cm$^{-1}$ (bending), 2971 cm$^{-1}$ (stretching) and 3031 cm$^{-1}$ (stretching) are attributed to the O-H bond vibrations. The source of these O-H vibrations could be the moisture absorbed from atmosphere by the samples.

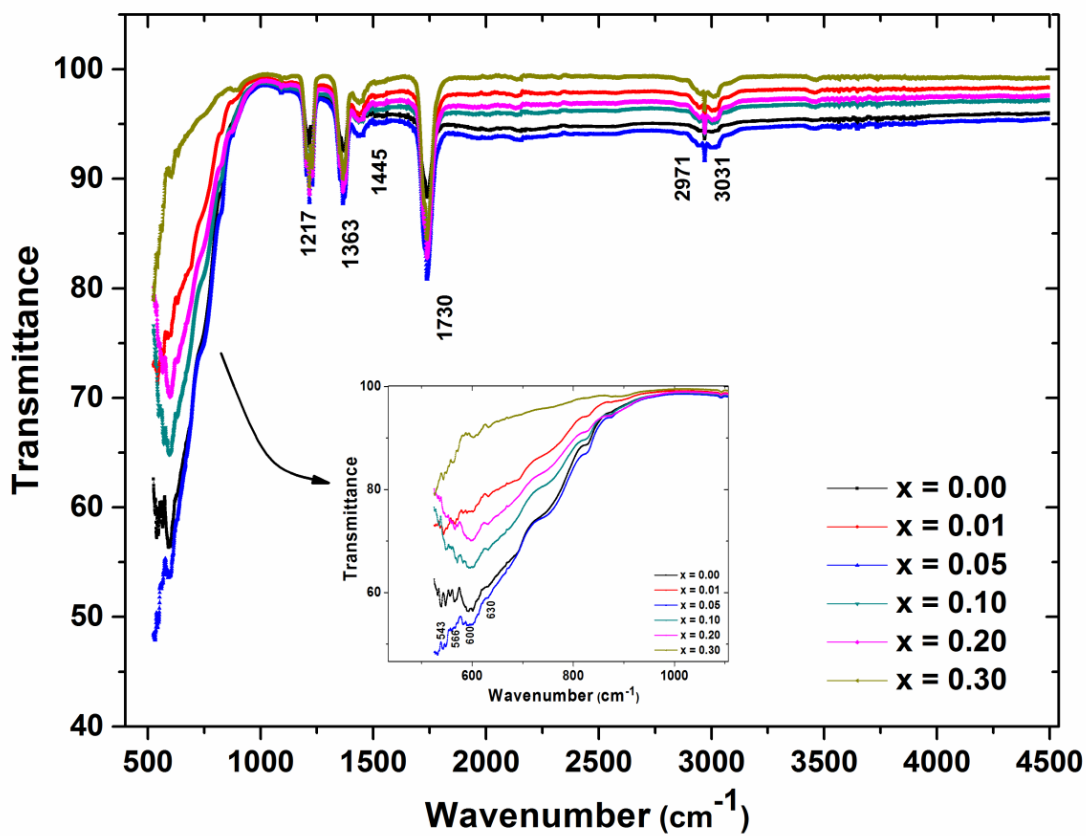

**Fig. 5. FTIR Spectra of Ba$_{1-x}$Sn$_x$Nb$_2$O$_6$ compositions**

## 3. AC Conductivity

The degree of covalence determining augmented dielectric nature of BSN compositions is verified using *ac* conductivity ($\sigma_{ac}$) investigations as proposed by Jonscher's universal power law[27]: $\sigma_{ac} = \sigma_{dc} + A\omega^n$, where, $\sigma_{dc}$ is the frequency independent *dc* conductivity contribution, A is a temperature dependent constant, $\omega$ is the angular frequency ($\omega = 2\pi f$) and *n* is the frequency exponent. The value of frequency exponent *n* determines the dominant conduction mechanism: $0<n<1$ associated with the hopping mechanism of ions and $n>1$ associated with the quantum mechanical tunneling of charge carriers over localized sites[28]. The conductivity values are calculated from the dielectric data using the relation $\sigma_{ac} = 2\pi f \varepsilon' \varepsilon_o \tan\delta$, where the symbols have their usual meaning. The variation of *ac* conductivity with frequency and temperature is investigated to conclude the nature of electronic transfer in the BSN samples, Fig.6. Note that *ac* conductivity values increase with frequency and tin concentration in BSN compositions. Typical values are 5.44µS/m (at 20°C and 50Hz), 67.6µS/m (at 20°C and 1 MHz) and 1028.5µS/m (at 300°C and 1 MHz). The values of *n* for all samples is computed using the slope on linear part of data curves in Fig.6, Table-3. The value of *n* increase with the increase in tin concentration as estimated earlier between the limits of hopping mechanism[28] confirming additionally *an increase in covalence* on doping tin. The increase in conductivity values with frequency could be due to lowering of the capacitive energy barrier around domain boundary. The *ac* activation is also probed by utilizing temperature dependence of dielectric data as shown in the inset of Fig.6 (In$\sigma_{ac}$ versus 1000/T). A typical insulating negative temperature coefficient of resistance (NTCR) behavior is observed for all BSN compositions. Near the phase transition temperatures, the rise in $\sigma_{ac}$ values of all the BSN samples is steep and is attributed to crystalline phase change from monoclinic to orthorhombic. The *ac* activation energy ($E_a$) is calculated using the slope of curves as plotted on ln$\sigma_{ac}$ vs. 1000/T graph[29] using the equation: $\sigma_{ac} = \sigma_0 \exp(-E_a/k_B T)$, where, $\sigma_0$ is a constant, $k_B$ is Boltzmann's constant and $T$ is absolute temperature. The activation energies of all samples are listed in Table-3.

**Table-3 Covalence parameter and *ac* activation energy for Ba$_{1-x}$Sn$_x$Nb$_2$O$_6$ compositions**

| x | n | $E_a$ (eV) |
|---|---|---|
| 0.00 | 0.24 | 0.258 |
| 0.01 | 0.26 | 0.261 |
| 0.05 | 0.30 | 0.265 |
| 0.10 | 0.37 | 0.375 |
| 0.20 | 0.42 | 0.239 |
| 0.30 | 0.49 | 0.198 |

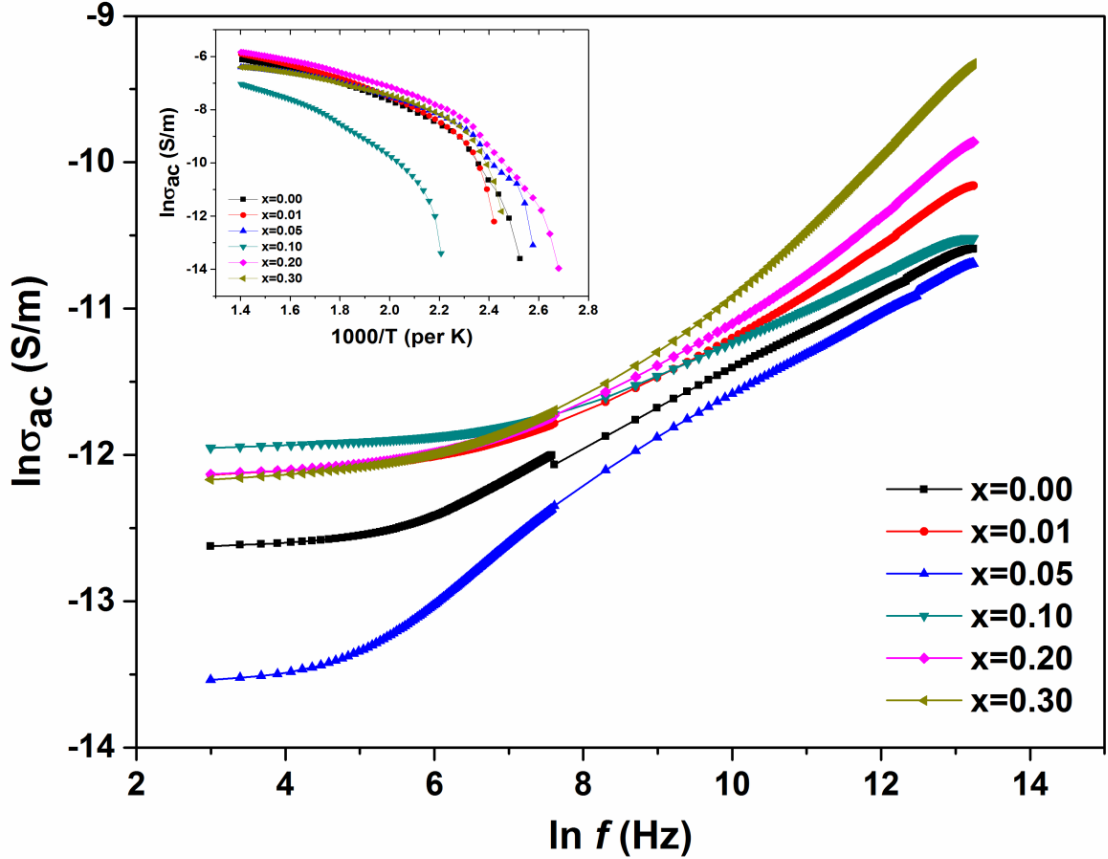

**Fig. 6. Variation of σ$_{ac}$ with frequency and temperature (inset)**

## 4. Hall Measurements

The measurements of absorption spectroscopy and microstruture study using TEM-SAED/HRTEM images in conjugation with dielectric investigations have indicated towards the following: a) tin doping beyond 5% generates *nano* regions of amorphous network as a result of metal alloy $Sn_2O_3$ introducing severe disorder in crystal structure of BN, b) overall low values of dielectric constant are useful in facilitating exciton seperation and c) transmittance to major visible spectrum. These indications prompted us to investigate BSN compositons for type of carrier concentration and hall resistivity using van der Pauw geometry[30]. The measurments are recorded on randomly selected BSN composition thin pellets (~0.5mm) at fixed magnetic field of 500 Gauss using four conducting electrodes on arbitrary thin pellet surface. The Hall voltage values were recorded across two diagonal contacts by decreasing current across other two diagonal contacts from 10 $m$A to 0.5$m$A. The data of Hall voltage (Fig.S5) was normalized according to expression, $V_H = (V_{12} + V_{21} + V_{34} + V_{43})/2$. The nature of carrier concentration in undoped BN was found to be $n$-type that was gradually converted into $p$-type on increasing tin doping up to 5%. An increase in tin doping beyond 10% follows a reverse trend. The sample with 10% tin doping possess $n$-type conduction presumably due to emerging metal clusters and contributing electrons as majority carriers. All BSN compositions were found to exhibit carrier concentration higher than $10^{20}$ per cm$^3$ with minimum Hall resistivity value ~ $3.4 \times 10^{-3}$ Ω-cm for 30% tin-doped BSN composition.

## IV. Conclusions

Single *monoclinic* phase barium tin niobate Ba$_{1-x}$Sn$_x$Nb$_2$O$_6$ BSN (x=0.00, 0.01, 0.05, 0.10, 0.20 and 0.30) ceramics have been successfully prepared using microwave synthesis. Tin occupied 15-coordinated A2-barium sites in all BSN compositions in addition to forming minor *locally distortive* monoclinic phase Sn$_2$O$_3$ metal alloy on high doping (≥10%). Tin doping introduced severe disorder in crystalline network promoting semiconducting nature of less strained dielectric BN compound. The reason for this disorder is the asymmetric electron distribution around tin sites due to well-known Sn 5s$^2$ sterically active lone pair. The asymmetric electron distribution is due to coupling between unfilled Sn (5p) with antibonding combination generated from interaction between Sn (5s) and O (2p) orbitals. These degenerate energy states affect intrinsic polarizability of atomic sites that is investigated through *raman* spectroscopy. The results of *uv-visible* absorption and *infrared* spectroscopy indicated towards significant changes in absorption edges and vibrational character of expected and new bonds. The values of absorbance, optical energy band gap, low dielectric constant and hall resistivity are in very close agreement with those reported for popular indium tin oxide (ITO). The investigated microwave synthesized BSN compositions offer uniqueness on many grounds like non-toxic, low cost, earth crust abundance and energy efficient method of preparation. This work proposed BSN compositions as a competent substitute for ITO.